\input{aipcheck}

\documentclass[
    ,final            
  ]
  {aipproc}

\layoutstyle{8x11single}

\usepackage{color}

\begin{document}

\quad{PI/UAN-2013-559FT}

\title{STATISTICAL ANISOTROPY IN INFLATIONARY MODELS WITH MANY VECTOR FIELDS AND/OR PROLONGED
ANISOTROPIC EXPANSION}


\author{L. Gabriel G\'omez}{
 address={Centro de Investigaciones en Ciencias B\'asicas y
Aplicadas, Universidad Antonio Nari\~no, \\
Cra 3 Este \#  47A - 15,
Bogot\'a D.C. 110231, Colombia.}
,altaddress={Escuela de F\'isica, Universidad Industrial de
Santander, \\
Ciudad Universitaria, Bucaramanga 680002, Colombia.}
 }
\author{Yeinzon Rodr\'iguez}{
address={Centro de Investigaciones en Ciencias B\'asicas y
Aplicadas, Universidad Antonio Nari\~no, \\
Cra 3 Este \#  47A - 15,
Bogot\'a D.C. 110231, Colombia.}
,altaddress={Escuela de F\'isica, Universidad Industrial de
Santander, \\
Ciudad Universitaria, Bucaramanga 680002, Colombia.}
 }

\begin{abstract}
We study the most general contributions due to scalar field
perturbations, vector field perturbations, and anisotropic
expansion to the generation of statistical anisotropy in the
primordial curvature perturbation $\zeta$. Such a study is done
using the $\delta N$ formalism where only linear terms are
considered. Here, we consider two specific cases that lead to
determine the power spectrum $P_{\zeta}(k)$ of the primordial curvature
perturbation. In the first one, we consider the
possibility that the $n$-point correlators of the field perturbations in real space are
invariant under rotations in space (statistical isotropy); as a
result, we obtain as many levels of statistical anisotropy as
vector fields present and, therefore, several preferred
directions. The second possibility arises when we consider
anisotropic expansion, which leads us to obtain $I+a$ additional
contributions to the generation of statistical anisotropy of
$\zeta$ compared with the former case, being $I$ and $a$ the number
of scalar and vector fields involved respectively.
\end{abstract}

\maketitle


\section{Introduction}
The most important quantity in modern cosmology is the primordial
curvature perturbation $\zeta$ since it is responsible for the
formation of structures such as galaxies and galactic clusters;
besides, $\zeta$ is directly related to the temperature fluctuations in the cosmic microwave background (CMB) through the
Sachs-Wolfe effect \cite{sach}. Hence, the statistical properties
of the CMB temperature anisotropies can be  described in terms of
the spectral functions, like the spectrum, bispectrum,
trispectrum, etc., of $\zeta$; in
this way, we may compare theory with observation. On the other
hand, we may see, from the distribution of temperatures in the CMB
\cite{Komatsu} and from the distribution of matter that form the
large-scale structure that our Universe exhibits,
departures from the exact homogeneity and isotropy. \\ \\
More precisely, recent observations of the CMB anisotropies in the
Wilkinson Microwave Anisotropy Probe (WMAP) experiment show that
there are certain features of the full sky maps which seem to be
anomalous \cite{groe,Eriksen1,Picon,Groeneboom,Eriksen}. These anomalies
indicate an alignment of the lowest multipoles in the quadrupole
moment of the power spectrum also known as
the ``axis of evil'' \cite{mae}, and an asymmetry in power between
the northern and southern ecliptic hemispheres (the hemispherical
anisotropy) \cite{Eriksen1,Eriksen}. These anomalies also indicate the presence of statistical anisotropy \cite{groe,Picon,Groeneboom} which
indicates one preferred direction in
the Universe; in other words, when we analyze the statistical
properties of the distribution of temperatures, the power spectrum
shows some dependence on the direction of the wavevector,
signaling in turn violation of the rotational invariance in the
two-point correlator of $\zeta$, a feature which is called
statistical anisotropy and is parameterized through the level of
statistical anisotropy $g_{\zeta}$ \cite{Ackerman}. However, it has been suggested
that systematic and statistical errors in the CMB signal may lead
to the observed anomalies since the preferred direction lies near
the plane of the solar system \cite{groe}. Moreover, the forthcoming
observations of the Planck satellite may as well detect
statistical anisotropy in the near future which would lead us to compare theory and observation more accurately \cite{planck,pullen}.\\ \\
The probability distribution function for $\zeta$ has well defined
statistical descriptors which depend directly upon the particular
inflationary model (once the action has been defined) and that are
suitable for comparison with present observational data. The right
framework to propagate the statistical properties of the field
perturbations to the statistical properties of $\zeta$ is the
cosmological perturbation theory (CPT) \cite{Weinberg}; however, this normally
involves lengthy calculations, even more when the nature of the
fields is not scalar. Despite of this, the CPT is valid throughout
all scales, leaving no room for discrepancies attributed to not
considered subhorizon phenomena. A different approach is the
$\delta N$ formalism, where $\zeta$ is identified with the perturbation in the amount of
expansion $N$ from an initial time in a flat slicing to a final
time in a uniform energy density slicing (the threading must be
comoving) \cite{staro,sasa,tanaka,malik}. The $\delta N$ formalism
gives an expression for $\zeta$ which is valid to all orders in
CPT; however, it is only valid for superhorizon scales (in
absolute contrast with CPT). Of course, extracting the statistical
properties of the distribution of $\zeta$ in the $\delta N$
formalism requires also to do some ``perturbation theory'': to
expand $N$ in a Taylor series and to cut it out at the desired
order.\\ \\
This paper is structured as follows. Initially we provide the
theoretical framework for building the $n$-point correlators of
$\zeta$. After that, we obtain the levels of statistical
anisotropy for $\zeta$ when isotropic (anisotropic) expansion and
multi-vector field perturbations are considered. When we consider
isotropic expansion only, one generic expression for $g_{\zeta}$ is
obtained; moreover there exist $a$ preferred directions in
contrast with the parameterization of a single vector field. Another
possibility arises when anisotropic expansion is present in the
background metric which leads to obtain $I+2a$
contributions to the generation of
statistical anisotropy in the primordial curvature perturbation.
Finally, we discuss our results and present our conclusions.

\section{Theoretical Framework}
We are  interested in studying the statistical properties of a
perturbation map through the $n$-point correlators of the
perturbations  \cite{Juanpa}. Let's define a scalar perturbation
$\beta(\textbf{x)}$ in real space, which can be expressed in terms
of a Fourier integral expansion as
\begin{equation}
\beta(\textbf{x})=\int \frac{d^{3}k}{(2\pi)^{3}} \beta(\textbf
{k})e^{i\textbf k.\textbf x}.\label{b2}
\end{equation}
The  $n$-point correlators of the perturbations
$\beta(\textbf{x})$  are defined as averages over the ensemble of
universes of the products  $\beta(\textbf{x}_{1})\beta(\textbf
{x}_{2}.)..\beta(\textbf {x}_{n})$, where ${\bf x}_{1},{\bf x}_{2},...,{\bf x}_{n}$
represent different points in space\footnote{ The ensemble average
inside the integral is over the Fourier mode functions only since
they are the stochastic variables.}:
\begin{equation}
\langle \beta(\textbf{x}_{1})
\beta(\textbf{x}_{2})...\beta(\textbf{x}_{n})\rangle = \int
\frac{d^3 k_{1} d^3 k_{2}} {(2\pi)^{3} (2\pi)^{3}}...\frac{
d^{3}k_{n}}{(2\pi)^{3}} \langle \beta(\textbf {k}_{1})
\beta(\textbf {k}_{2})...\beta\textbf{k}_{n}\rangle e^{i(\textbf
k_{1}.\textbf x_{1}+\textbf k_{2}.\textbf x_{2}...+\textbf
k_{n}.\textbf x_{n})}\label{b3}.
\end{equation}
Hence, the correlation functions in real space may be studied via
the correlation functions in momentum space. Let's see now the
meaning of statistical homogeneity and statistical isotropy in the
$n$-point correlators of perturbations.

\subsection{Statistical homogeneity}

According to the observation, the perturbation map is not
homogeneous, but it may be that the probability distribution
function governing $\beta(\textbf{x)}$ is. This feature can be
expressed in the following way: if the $n$-point correlators in
real space are invariant under translations in space, it is said
that there exists statistical homogeneity \cite{Konstan,Abramo}
\begin{equation} \langle \beta(\textbf x_{1}+\textbf{d}) \beta(\textbf
x_{2}+\textbf{d})...\beta(\textbf x_{n}+\textbf{d})\rangle=\langle
\beta(\textbf x_{1}) \beta(\textbf x_{2})...\beta(\textbf
x_{n})\rangle,\label{sh}
\end{equation}
where $\textbf{d}$ is some vector in real space establishing the
amount of spatial translation. The above condition may be achieved
if the $n$-point correlators in momentum space are proportional to
a Dirac delta function (which is not the only possibility).
\begin{equation} \langle \beta(\textbf k_{1}) \beta(\textbf
k_{2})...\beta(\textbf k_{n})\rangle\equiv(2\pi)^{3}\delta ^{3}
(\textbf k_{1}+\textbf k_{2}+...+\textbf k_{n})M_{\beta}(\textbf
k_{1},\textbf k_{2},...,\textbf k_{n}),
\label{a1122}\end{equation} where the function $M_{\beta}(\textbf
k_{1},\textbf k_{2},...,\textbf k_{n})$ is called the
$(n-1)$-spectrum. Statistical homogeneity is absolutely necessary
as an hypothesis of the ergodic theorem \cite{Weinberg}.

\subsection{Statistical isotropy}

Once statistical homogeneity has been secured, we ask about the
invariance under spatial rotations of the $n$-point correlators in
real space (i.e. statistical isotropy). Of course again, the perturbation map
is not isotropic, but it may be that the probability distribution
function governing $\beta({\bf x})$ is, which is called statistical
isotropy \cite{Konstan,Abramo}. This means that the $n$-point
correlators in real space are invariant under rotations in space.
\begin{equation} \langle \beta(\tilde{\textbf{x}}_{1})
\beta(\tilde{\textbf{x}}_{2})...\beta(\tilde{\textbf{x}}_{n})\rangle=
\langle \beta(\textbf{x}_{1})
\beta(\textbf{x}_{2})...\beta(\textbf{x}_{n})\rangle,\label{si}\end{equation}
where $\tilde{\textbf{x}}_{i} = R \textbf{x}_{i}$, $\emph{R}$
being a rotation operator. To satisfy the above requirement, the
$(n-1)$-spectrum must satisfy the condition
\begin{equation}
M_{\beta}(\tilde{\textbf k}_{1},\tilde{\textbf
k}_{2},...,\tilde{\textbf k}_{n})=M_{\beta}(\textbf k_{1},\textbf
k_{2},...,\textbf k_{n}),\label{b0}\end{equation} where the tildes
over the momenta represent as well a spatial rotation,
parameterized by $\emph{R}$, in momentum space. We assume that the
$n$-point correlators are invariant under translations in space
(statistical homogeneity):
\begin{equation}
\langle\beta(\textbf{k}_{1})\beta({\textbf{k}_{2}})\rangle\equiv(2\pi)^{3}\delta^{3}(\textbf
k_{1} + \textbf k_{2} ) P_{\beta}({\bf k}_1).\label{b4}
\end{equation}
In the above expressions, $P_{\beta}$, is called the spectrum.
When we consider statistical isotropy, the spectrum depends only
on the wavenumber. Hence, the argument in the spectrum may be
considered as $k = |k_{1}| = |k_{2}|$. \\ \\
Considering   violation of rotational invariance due to the
presence of  a vector field which points out in the preferred
direction $\hat{\textbf{d}}$,  the  form of the power spectrum
changes according to \cite{Ackerman}
\begin{equation}
P_{\zeta}(\textbf{k})=P^{iso}_{\zeta}(k)[1+g_{\zeta}(\hat{\textbf{k}} \cdot \hat{\textbf{d}})^{2}].\label{b1}
\end{equation}
In the above expression $P^{iso}_{\zeta}(k)$ is the average over
all directions, $\hat{\textbf{k}}$ is a unit vector and
$g_{\zeta}$ is the level of statistical anisotropy whose value is
in the range $g_{\zeta}=0.290\pm 0.031$ and rules out statistical
isotropy at more than $9\sigma$ \cite{groe}.

\section{The $\delta N$ formalism with many vector and scalar fields}

Starting with an initial flat slicing such that the
locally-defined scale factor is homogeneous, and ending with a
slicing of uniform energy density, we can then express the primordial
curvature perturbation $\zeta$ through the $\delta N$ formula
\cite{Lyth}
\begin{equation}
\zeta(\textbf{x},t)=N(\textbf{x},t)-N_{0}(t)=\delta
N(\textbf{x},t) \label{a5}, \end{equation} where $N$ is the associated amount of expansion. Thus, the evolution of
$\zeta$ depends directly on the evolution of $N$ and this in turn
depends on the nature of the fields considered. In
this work, we will consider $n$ scalar fields and $m$ vector
fields. To deal with the different contributions from the fields
involved, we introduce the notation:
\begin{equation} \delta\Phi_{A}\equiv\{\delta\phi_{I},\delta
A_{i}^{a}\}.\end{equation} The index $A$  is separated into two
sets: a set of indices  $I$ labelling the scalar fields which run
from 1 to $n$ and another set of indices $a$ labelling vector
fields which run from 1 to $m$. The index $i$ specifies the
component of any vector field and runs from 1 to 3.
Accordingly, the derivatives of $N$ with respect to the fields are
separated as follows \cite{Juanpa}
\begin{equation}
N_{A}=\{N_{I}\;,N_{j}^{a}\},\end{equation}
\begin{equation} N_{AB}=\{N_{IJ}\;,N_{Ij}^{b}\;,N_{ij}^{ab}\}.
\end{equation}
In the above notation, we represent the mixed second derivative
with respect to $\phi$  and $A_{j}^{b}$  as $N_{Ij}^{b}=\partial
N/\partial \phi_{I}A_{j}^{b}$. In terms of the  above introduced
notation, the curvature perturbation from the multi-scalar and
multi-vector field case is written in terms of the mode functions
by means of the following truncated expansion up to first order:
\begin{equation}
\zeta(\textbf{k},t)=N_{A}\delta\Phi(\textbf{k},t)+\frac{1}{2}N_{AB}\int
\frac{d^{3}k_{1}}{(2\pi)^{3}}\delta\Phi_{A}(\textbf{k}-\textbf{k}_{1})\delta\Phi_{B}(\textbf{k}_{1}).\label{zeta}
\end{equation}
We only consider linear terms of $\zeta$ in Eq. (\ref{zeta})
(corresponding to what we call the tree level contributions)
because we  are interested in studying  the power spectrum of
$\zeta$ and not in other spectral functions. Hence, once the
product of the respective perturbations has been made, we perform
the average over the ensemble in order to obtain the two-point
correlator associated to $\zeta$
\begin{eqnarray}
\langle\zeta(\textbf{k}_{1})\zeta(\textbf{k}_{2})\rangle&=&N_{A}N_{B}
\langle\delta\Phi_{A}(\textbf{k}_{1})\delta\Phi_{B}(\textbf{k}_{2})\rangle\\
&=&N_{\phi}^{2}\langle\delta\phi(\textbf{k}_{1})\delta\phi(\textbf{k}_{2})\rangle
+N_{i}N_{j}\langle\delta A_{i}(\textbf{k}_{1})\delta
A_{j}(\textbf{k}_{2})\rangle \nonumber\\
&+& N_{\phi}N_{j}\langle\delta\phi(\textbf{k}_{1})\delta
A_{j}(\textbf{k}_{2})\rangle +N_{i} N_{\phi}\langle\delta
A_{i}(\textbf{k}_{1})\delta\phi(\textbf{k}_{2})\rangle.\label{aa1}
\end{eqnarray}
We can see that the two-point correlator of $\zeta$ depends on the
two-point correlators of the involved fields, (either scalar,
vector, or in the most general case both fields). In view of this
fact, it is necessary to assume some properties of these
correlators; hence, we will assume the following preposition as a
conjecture to obtain the power spectrum of the curvature
perturbation and therefore the levels of statistical anisotropy \cite{Gomez}:
if we consider anisotropic expansion (the field
perturbations live in the anisotropic background metric), at
least one of the correlators of the involved field perturbations is not invariant under rotations in space; it is  reasonable to
think this since there is no a conceivable way to get statistical
isotropy if the perturbations are defined in an anisotropic
background. Let's consider anisotropic  expansion so that the two-point correlator of $\zeta$ is  not invariant under rotations in
space; however, it will be absolutely necessary to assume
statistical homogeneity as an hypothesis of the ergodic theorem.
These features lead to express the two-point correlator of the field perturbations
as
\begin{equation}
\langle\delta\Phi_{A}(\textbf{k}_{1})\delta\Phi_{B}(\textbf{k}_{2})\rangle=(2\pi)^{3}\delta(\textbf{k}_{1}+\textbf{k}_{2})\Pi_{AB}(\textbf{k}_{1}),
\end{equation}
where the power spectra of the scalar, vector and mixed fields
perturbation are
$\Pi_{AB}=\{\Pi_{IJ}(\textbf{k}),\Pi_{Ij}^{b}(\textbf{k}),\Pi_{iJ}^{a}(\textbf{k}),\Pi_{ij}^{ab}(\textbf{k})\}$
respectively. Thus, the power spectrum of $\zeta$ is determined by
the power spectra of the field perturbations and the derivatives of the amount of expansion
$N_{A}$ which, in turn, depend on the background metric:
\begin{equation}
P_{\zeta}(\textbf{k}_{1})=N_{A}N_{B}\Pi_{AB}(\textbf{k}_{1}).\label{b8}
\end{equation}
In addition, the power spectrum of each field could be different
since each one of the associated spectra exhibits some dependence
on the wavevectors due to the lack of symmetry in the background
and therefore in the $n$-point correlators.  These reasons lead us to
have an anisotropic spectrum for $\zeta$ sourced by the statistical anisotropy in the two-point correlator of the field perturbations.\\ \\
On the other hand, the scalar  field perturbation power spectra
$\Pi_{IJ}(k)$, related to the two-point correlators of the scalar
fields $\phi_{I}$, is defined as
\begin{equation}
\langle\delta\phi_{I}(\textbf{k}_{1})\delta\phi_{J}(\textbf{k}_{2})\rangle=(2\pi)^{3}\delta(\textbf{k}_{1}+\textbf{k}_{2})\Pi_{IJ}(\textbf{k}_{1}),
\end{equation}
whereas for the vector field perturbation power spectra
$\Pi_{ij}^{ab}$, whose origin lies again in the anisotropic
expansion, we have\footnote{The scalar-vector field perturbation power
spectra  do not contribute to  the power spectrum of $\zeta$
because there is no correlation between  fields of different
nature.}
\begin{equation} \langle\delta A_{i}^{a}(\textbf{k}_{1})\delta
A_{j}^{b}(\textbf{k}_{2})\rangle=(2\pi)^{3}\delta^{3}(\textbf{k}_{1}+\textbf{k}_{2})\Pi_{ij}^{ab}(\textbf{k}_{1}).\end{equation}
According to Ref. \cite{Konstan},  $\Pi_{ij}^{ab}$ is
given by
\begin{equation}
\Pi_{ij}^{ab}(\textbf{k})=\Pi_{ij}^{even}(\textbf{k})P_{+}^{ab}(\textbf{k})+i\Pi_{ij}^{odd}(\textbf{k})P_{-}^{ab}(\textbf{k})
+\Pi_{ij}^{long}(\textbf{k})P_{long}^{ab}(\textbf{k}),\label{b8c}\end{equation}
which is written in terms of  the longitudinal component of the
power spectra $P_{long}^{ab}$
 and the parity-conserving and parity-violating power spectra $P_{+}^{ab}$  and $P_{-}^{ab}$  respectively. These  spectra are defined as
\begin{equation}
P_{\pm}^{ab}\equiv\frac{1}{2}(P_{R}^{ab}+P_{L}^{ab}),\end{equation}
where $P_{R}^{ab}$  and $P_{L}^{ab}$ denote the power spectra for
the transverse components with right-handed and left-handed
circular polarizations, and again all of them depend on the
wavevector because we are considering anisotropic expansion.  The
formal definitions of the polarization spectra are given by
\begin{equation} \langle\delta
A_{\lambda}^{a}(\textbf{k}_{1})\delta
A_{\lambda}^{b\ast}(\textbf{k}_{2})\rangle=(2\pi)^{3}\delta^{3}(\textbf{k}_{1}-\textbf{k}_{2})P_{\lambda}^{ab}(\textbf{k}_{1}),
\end{equation}
where we have used the reality condition
$\beta(-\textbf{k})=\beta^{*}(\textbf{k})$ and $\lambda$ denotes
the different polarizations $ L$, $R$, or $long$. The basis
$\Pi^{even}_{ij}(\textbf{k})$, $\Pi^{odd}_{ij}(\textbf{k})$ and
$\Pi^{long}_{ij}(\textbf{k})$ in this case is given by
 \begin{equation}
\Pi^{even}_{ij}(\textbf{k})\equiv
\delta_{ij}-\hat{k}_{i}\hat{k}_{j},\;\;\;\Pi^{odd}_{ij}(\textbf{k})\equiv
\epsilon_{ijk}\hat{k_{k}},\;\;\;\Pi^{long}_{ij}(\textbf{k})\equiv
\hat{k}_{i}\hat{k}_{j},\label{b9}\end{equation} where
$\epsilon_{ijk}$ is the totally antisymmetric tensor. Now, it is
possible to calculate the power spectrum of  $\zeta$ in terms of
the components associated to the scalar and  vector field
perturbation power spectra defined respectively above. Starting
from  Eq. (\ref{b8}), we can obtain the remarkable result
\cite{Gomez}
\begin{eqnarray}
P_{\zeta}(\textbf{k}_{1})&=&(N_{\phi}^{I})^{2}P_{\delta\phi}^{I}(\textbf{k}_{1})
+N_{i}^{a}N_{j}^{b}[(\delta_{ij}-\hat{k}_{1i}\hat{k}_{1j})P_{+}^{ab}(\textbf{k}_{1})\delta_{ab}
+i\epsilon_{ijk}\hat{k}_{1k}P_{-}^{ab}(\textbf{k}_{1})\delta_{ab} \nonumber \\
&+&\hat{k}_{1i}\hat{k}_{1j}P_{long}^{ab}(\textbf{k}_{1})\delta_{ab}].\end{eqnarray}
In this expression, we have not considered correlation between
different scalar field perturbations and different vector field
perturbations; however, it is the most general expression for
$P_{\zeta}$ since parity violation in the Lagrangian has been
initially allowed (although, this term really does not contribute to $P_{\zeta}$ due to symmetry arguments) and the anisotropic
expansion has been considered. Besides, we have considered the
possibility of having several scalar and vector fields in the
inflationary dynamics.

\subsubsection{Case I. Isotropic expansion}
 As a particular case, when isotropic expansion
is considered (which leads to have invariance under rotations in
space for  the two-point correlators of the fields involved)  we
obtain a similar expression for the power spectrum of $\zeta$:
\begin{equation}
P_{\zeta}(\textbf{k}_{1})=[(N_{\phi}^{I})^{2}P_{\delta\phi}^{I}(k_{1})+
N_{i}^{a2}P_{+}^{a}(K_{1})]+(\hat{N}^{a}\cdot\hat{k}_{1})^{2}[P_{long}^{a}(k_{1})-P_{+}^{a}(k_{1})],\label{b11}
\end{equation}
and hence, we obtain $a$ levels of statistical anisotropy:
\begin{equation}
g_\zeta^{a}=\frac{(r_{long}^{a}-1)(N_{i}^{a})^{2}P_{+}^{a}(k_{1})}{(N_{\phi}^{I})^{2}P_{\delta\phi}^{I}(k_{1})+(N_{i}^{a})^{2}P_{+}^{a}(k_{1})}.\label{b12}
\end{equation}
The latest result can be compared with that obtained in Ref. \cite{Konstan}.
Moreover, in this case there are as many preferred directions as
the number of vector fields present; namely, there now exist $a$
preferred directions or levels of statistical anisotropy in
contrast with the parameterization given by Eq. (\ref{b1}). In
the above expression, $P_{\zeta}^{iso}$ has been identified  as
\begin{equation}
P_{\zeta}^{iso}=(N_{\phi}^{I})^{2}P_{\delta\phi}^{I}(k_{1})+(N_{i}^{a})^{2}P_{+}^{a}(k_{1}),\label{b13}\end{equation}
and we have defined $\textbf{N}^{a}\equiv
N_{i}^{a}\hat{\textbf{N}}$ and $r_{long}=P_{long}/P_{+}$.

\subsubsection{Case II. Anisotropic expansion}
The most interesting case arises when anisotropic expansion is
assumed which is equivalent to allow violation of the statistical
isotropy in the two-point correlators of the field perturbations whose
associated spectra are $P_{\delta\phi}$, $P_{+}$, $P_{-}$, and $P_{long}$
respectively. Considering the parameterization given by Eq.
(\ref{b1}) and linear contributions only, the power spectrum of
$\zeta$ may now be written as \cite{Gomez}
\begin{equation}
P_{\zeta}(\textbf{k}_{1})=P_{\zeta}^{iso}(k_{1})[1+\tilde{g}_{\delta\phi}^{I}(\hat{\textbf{k}}_{1}\cdot
\hat{\textbf{d}}_{\delta\phi}^{I})^{2}+\tilde{g}_{+}^{a}(\hat{\textbf{k}}_{1}\cdot\hat{\textbf{d}}_{+}^{a})^{2}+
\tilde{g}_{N}^{a}(N_{i}^{a})^{2}(\hat{\textbf{k}}_{1}\cdot\hat{\textbf{N}}^{a}_{1})^{2}].\label{b14}\end{equation}
This result clearly exhibits the nature of the origin of $2a+I$
preferred directions determined by
$\hat{\textbf{d}}_{\delta\phi}^{I}$, $\hat{\textbf{d}}_{+}^{a}$
and $\hat{\textbf{N}}^{a}_{1}$. The levels of statistical
anisotropy $\tilde{g}_{N}^{a}$ in Eq. (\ref{b14}) were
precisely obtained in the isotropic expansion case; however, in the present case,
$N$ depends on the anisotropic background. Furthermore, although
there is no explicit dependence for the scalar field on the
wavevector, the anisotropic contribution of these fields is due to
the fact that we are considering anisotropic expansion. The levels
of statistical anisotropy associated with each field are given by
\begin{equation}
\tilde{g}_{\delta\phi}^{I}=\frac{g_{\delta\phi}^{I}(N_{\phi}^{I})^{2}P_{\delta\phi}^{I\;iso}(k_{1})}
{(N_{\phi}^{I})^{2}P_{\delta\phi}(k_{1})+(N_{i}^{a})^{2}P_{+}^{a}(k_{1})},\label{b15}
\end{equation} \begin{equation}
\tilde{g}_{+}^{a}=\frac{g_{+}^{a}(N_{i}^{a})^{2}P_{+}^{a\;iso}(k_{1})}
{(N_{\phi}^{I})^{2}P_{\delta\phi}(k_{1})+(N_{i}^{a})^{2}P_{+}^{a}(k_{1})},\label{b16}
\end{equation} \begin{equation}
\tilde{g}_{N}^{a}=\frac{(r_{long}^{a}-1)(N_{i}^{a})^{2}P_{+}^{a\;iso}(k_{1})}
{(N_{\phi}^{I})^{2}P_{\delta\phi}(k_{1})+(N_{i}^{a})^{2}P_{+}^{a}(k_{1})}.\label{b17}
\end{equation}
In these expressions, the isotropic spectra do not correspond to
the isotropic case because the nature of the background metric
is now anisotropic. Rather than calculating these spectra in an
isotropic background, which is incoherent, we will refer to them
as the pieces of the spectra that
do not depend on the wavevector.\\ \\
As a remarkable result, we have considered the most general case
to the generation of statistical anisotropy in the primordial
curvature perturbation $\zeta$ (and therefore to the power
spectrum in Eq. (\ref{b14})) due to the contributions of multiple field
perturbations and anisotropic expansion. This latter effect, is
clearly indicated by the $2a+I$ contributions. On the other
hand, in order to obtain the magnitude of the levels of
statistical anisotropy  when one particular setup is established (in
this new scenario), it will be necessary to study the anisotropic
particle production for the scalar and vector fields involved and to calculate
the amount of expansion in an anisotropic background; in this way,
we may directly compare our theoretical predictions with
observations.


\begin{theacknowledgments}
We would like to thank the organizers of the IX Mexican School of the DGFM-SMF for his hospitality and attention. L. G. G. acknowledges support for mobility from VIE (UIS) grant number 2012007978. Y. R. acknowledges support for mobility from VCTI (UAN). L. G. G. and Y. R. are supported by Fundaci\'on para la Promoci\'on de
la Investigaci\'on y la Tecnolog\'ia del Banco de la Rep\'ublica
(COLOMBIA) grant number 3025 CT-2012-02. In addition, Y. R. is
supported by VCTI (UAN) grant number 2011254.
\end{theacknowledgments}




\bibliographystyle{aipproc}   





\end{document}